\documentclass[pra,twocolumn,showpacs,preprintnumbers,superscriptaddress]
{revtex4}

\usepackage{times}
\usepackage{bm}
\usepackage{graphicx}
\usepackage{amsbsy}
\usepackage{amsmath}
\usepackage{amsfonts}
\usepackage{amsthm}

\begin{document}

\theoremstyle{plain}
\newtheorem{theorem}{Theorem}
\newtheorem{lemma}[theorem]{Lemma}
\newtheorem{corollary}[theorem]{Corollary}
\newtheorem{proposition}[theorem]{Proposition}
\newtheorem{conjecture}[theorem]{Conjecture}

\theoremstyle{definition}
\newtheorem{definition}[theorem]{Definition}

\theoremstyle{remark}
\newtheorem*{remark}{Remark}
\newtheorem{example}{Example}

\title{Usefulness of  Multiqubit W-type States in Quantum Information Processing Task}

\author{Parvinder Singh}
\thanks{psdcuraj.cs@gmail.com}
\affiliation{Indian Institute of Technology Jodhpur,
Rajasthan-342011, India}
\author{Satyabrata Adhikari}
\thanks{tapisatya@gmail.com}
\affiliation{Institute of Physics, Bhubaneswar-751005, India}
\author{Atul Kumar}
\thanks{atulk@iitj.ac.in}
\affiliation{Indian Institute of Technology Jodhpur,
Rajasthan-342011, India}
\date{\today}

\begin{abstract}
We analyze the efficacy of multiqubit W-type states as resources for quantum information. For this, we identify and generalize
four-qubit W-type states. Our results show that the states can be used as resources for deterministic quantum information processing.
The utility of results, however, is limited by the availability of experimental setups to perform and distinguish multiqubit measurements.
We, therefore, emphasize on another protocol where two users want to establish an optimal bipartite entanglement using the partially
entangled W-type states. We found that for such practical purposes, four-qubit W states can be a better resource in comparison to
three-qubit W-type states. For dense coding protocol, our states can be used deterministically  to send two bits of classical message
by locally manipulating a single qubit.
\end{abstract}


\maketitle

\section{Introduction}
Quantum entanglement \cite{Einstein} plays a key role in many potential applications in quantum information and computation  \cite{Bennett1,Bennett2,Bostrom,Gisin,Zukowski}. The optimal success of a quantum communication protocol can be ascertained by use of maximally entangled states as resources for information transfer. However, in general, the use of nonmaximally entangled resources leads to probabilistic protocols and the fidelity of information transfer is always less than unity. For example, quantum teleportation of a single qubit using a three and four-qubit W state is always probabilistic and teleportation fidelity depends on the unknown parameter of the teleported state. On the other hand, Agrawal and Pati \cite{Pati1} proposed a new class of three-qubit W-type states for deterministic teleportation of a single qubit by performing three-qubit joint measurements. The efficiency of these W-type states, however, decreases if one performs standard two-qubit and single qubit measurements only\cite{Adhikari3} instead of performing a joint three-qubit measurement. We address the question of usefulness of such non-maximally entangled  resources for sending maximum information from a sender to a receiver. \par
We propose a new class of non-maximally entangled four-qubit W-type states for quantum information processing and demonstrate the possibility of deterministic teleportation of a single qubit with unit fidelity. For practical purposes, we emphasize on a protocol to share optimal bipartite entanglement. For this, we use partially entangled four-qubit W-type states as a starting resource between the two users and achieve the optimal bipartite entanglement by performing standard two-qubit measurements only.
Our results show that the shared two qubit entanglement can lead to a maximally entangled resource for certain state parameters.
We further demonstrate the need to analyze four-qubit W-type states by comparing the efficacy of three and four qubit W-type states as resources in terms of concurrence \cite{Wootters} of the finally shared entangled state between the two users. Interestingly, our
results show that for certain ranges of parameters, four qubit W-type states are more efficient resources in comparison to three qubit W-type states for achieving optimal concurrence. \par
For dense coding, we found that in principle a sender can transmit 2-bit classical message to a receiver by locally manipulating his/her single qubit. The teleportation and dense coding protocols are also generalized for N-qubit W-type states.
\section{Telportation Using 4-particle W-type State }
Teleportation is a process to transmit quantum information over arbitrary distances using a
shared entangled resource. Although non-maximally entangled four-qubit W states can be used as
resources for probabilistic teleportation of a single qubit \cite{Shi}, one cannot achieve teleportation
of a single qubit using W states with certainty. We propose a new class of
four qubit W states, namely
\begin{eqnarray}
\left|\Psi_{k}\right\rangle_{1234} &=&
\frac{1}{2\sqrt{k+1}}\left[\left|1000\right\rangle+\sqrt{k}e^{i\gamma}\left|0100\right\rangle\right.
\nonumber \\ &+&\left.
\sqrt{k+1}e^{i\delta}\left|0010\right\rangle \right. + \left.
\sqrt{2k+2}e^{i\zeta}\left|0001\right\rangle\right]_{1234}
\end{eqnarray}
that can be used for deterministic quantum teleportation. For example, if Alice wants to teleport an unknown state
$\left|\phi\right\rangle_{a}=\left[\alpha\left|0\right\rangle+\beta\left|1\right\rangle\right]_{a}~,~\alpha^{2}+\beta^{2}=1$ to Bob, then Alice and Bob need to share the four qubit state $\left|\Psi_{k}\right\rangle_{1234}$ such that  Alice has qubits $1$, $2$ and $3$ and Bob has qubit $4$. In Eq. (1), $k$ is a real number and $\gamma,\delta,\zeta$ represent phases. \par
The joint state of five qubits can be represented as
\begin{eqnarray}
\left|\Phi\right\rangle_{a1234} &=& \left|\phi\right\rangle_{a} \otimes \left|\Psi_{k}\right\rangle_{1234}
\end{eqnarray}
In order to teleport the unknown state to Bob, Alice projects her four qubits on the states
\begin{eqnarray}
\left|\eta_{k}\right\rangle^{\pm}_{a123} &=& \frac{1}{2\sqrt{k+1}}
\left[\left|0100\right\rangle+\sqrt{k}e^{i\gamma}\left|0010\right\rangle\right.
\nonumber \\ &\pm&
\left.\sqrt{k+1}e^{i\delta}\left|0001\right\rangle \right. \pm
\left. \sqrt{2k+2}e^{i\zeta}\left|1000\right\rangle
\right]_{a123}\nonumber \\ \left|\xi_{k}\right\rangle^{\pm}_{a123}
&=& \frac{1}{2\sqrt{k+1}}
\left[\left|1100\right\rangle+\sqrt{k}e^{i\gamma}\left|1010\right\rangle
\right. \nonumber \\ &\pm&
\left.\sqrt{k+1}e^{i\delta}\left|1001\right\rangle\right. \pm
\left.\sqrt{2k+2}e^{i\zeta}\left|0000\right\rangle\right]_{a123}
\end{eqnarray}
Although the teleportation protocol works for all $k$, $\gamma$, $\delta$  and $\zeta$, for
simplicity we assume $k=1$ and $\gamma$ $=$ $\delta$ $=$ $\zeta$ $=0$. Thus, the joint state of
five qubits can be re-expressed using Alice's measurement basis as
\begin{eqnarray}
\left|\Phi\right\rangle_{a1234} &=&
\frac{1}{2}\left[\left|\eta_{1}\right\rangle^{+}_{a123}\left|\phi\right\rangle_{4}
+
\left|\eta_{1}^{-}\right\rangle^{+}_{a123}\sigma_{z}\left|\phi\right\rangle_{4}
\right. \nonumber \\ &+&
\left.\left|\xi_{1}^{+}\right\rangle^{+}_{a123}\sigma_{x}\left|\phi\right\rangle_{4}
\right. +  \left.
\left|\xi_{1}^{-}\right\rangle^{+}_{a123}\iota\sigma_{y}\left|\phi\right\rangle_{4}\right]
\end{eqnarray}
where $\left|\phi\right\rangle_{4}=[\alpha\left|0\right\rangle +\beta\left|1\right\rangle]_{4}~,~\alpha^{2}+\beta^{2}=1 $.\\
A four-qubit joint measurement on qubits $a, 1, 2$ and $3$ will project the
state of Bob's qubit onto one of the four possible states as shown in Eq. (4) with the equal probability of 1/4. \par
Hence, teleportation of a single qubit using non maximally entangled four qubit W state is
always successful. The use of proposed states as quantum channels also provides flexibility to the experimental set-ups by relaxing the
requirement of a maximally entangled shared resource for a faithful teleportation. Since the teleportation is deterministic,
the total probability and fidelity of teleporting a single qubit using a partially entangled
four-qubit W state is also unity.
\section{Teleportation Using W-type State of n-Particle System}
In the previous section, we have successfully demonstrated the
efficient quantum teleportation of a single qubit state using a
new class of four-qubit W-type state. We now extend our idea to
$n$-particle W-type states. \par In order to teleport the single
qubit state $\left|\phi\right\rangle_{a}$ to Bob, Alice needs to
share a $n$-particle state
\begin{eqnarray}
 \lefteqn{\left|\Psi_{k}\right\rangle_{12..n} \nonumber} && \nonumber \\ && \frac{1}{\sqrt{(n-2)(2k+n-3)+2}} \left[\left|100...n\right\rangle_{12...n}\right. \nonumber \\ &+&  \left.\sqrt{k}e^{i\gamma}\left|010...n\right\rangle_{12..n}+\sqrt{k+1}e^{i\delta}\left|001...n\right\rangle_{12..n}\right. \nonumber \\ &+& \left....\sqrt{k+(n-3)}e^{i\zeta}\left|000...10\right\rangle_{12..n}\right. \nonumber \\ &+& \left. \sqrt{(n-2)k+\frac{(n-2)(n-3)}{2}+1}e^{i\beta}\right. \nonumber \\ & & \left.\left|000...1\right\rangle_{12...n}\right]
\end{eqnarray}
with Bob such that particles $1$ to $n-1$ are with Alice and particle $n$ is with Bob. In this case, the projection bases used by Alice are
\begin{eqnarray}
\lefteqn{\left|\eta_{k}\right\rangle^{\pm}_{a,1,2...,n-1} =} &&\nonumber \\ && \frac{1}{\sqrt{(n-2)(2k+n-3)+2}} \left[\left|010...n\right\rangle\right. \nonumber \\ &+& \left. \sqrt{k}e^{i\gamma}\left|001...n\right\rangle + \sqrt{k+1}e^{i\delta}\left|0001...n\right\rangle \right. \nonumber \\ &+& \left. .....\sqrt{k+(n-3)}e^{i\zeta}\left|000...1\right\rangle\right.\nonumber \\ &\pm & \left.\sqrt{(n-2)k+\frac{(n-2)(n-3)}{2}+1}e^{i\beta}\right. \nonumber \\ & & \left.\left|100...0\right\rangle\right]_{a,1,2...,n-1}
\end{eqnarray}
\begin{eqnarray}
\lefteqn{\left|\xi_{k}\right\rangle^{\pm}_{a,1,2...,n-1}=} && \nonumber \\ && \frac{1}{\sqrt{(n-2)(2k+n-3)+2}}\left[\left|110...n\right\rangle\right. \nonumber \\ &+&  \left.\sqrt{k}e^{i\gamma}\left|101...n\right\rangle + \sqrt{k+1}e^{i\delta}\left|1001...n\right\rangle \right. \nonumber \\ &+&  \left. ....\sqrt{k+(n-3)}e^{i\zeta}\left|100...1\right\rangle\right. \nonumber \\ &\pm & \left. \sqrt{(n-2)k+\frac{(n-2)(n-3)}{2}+1}e^{i\beta}\right. \nonumber \\ & &  \left.\left|000...0\right\rangle\right]_{a,1,2...,n-1}
\end{eqnarray}
Similar to the teleportation protocol discussed in the previous section, we can express the joint state of $n+1$ particles in terms of
Alice's projection bases as
\begin{eqnarray}
\left|\Phi\right\rangle_{a12...n}&=&\left|\phi\right\rangle_{a}\otimes\left|\Psi_{k}\right\rangle_{123...n} \nonumber \\
&=&\frac{1}{2}\left[\left|\eta_{k}\right\rangle^{+}_{a12...n-1}\left|\phi\right\rangle_{n}\right. \nonumber \\ &+ &  \left. \left|\eta_{k}\right\rangle^{-}_{a12...n-1}\sigma_{z}\left|\phi\right\rangle_{n}\right. \nonumber \\ &+& \left. \left|\xi_{k}\right\rangle^{+}_{a12...n-1}\sigma_{x}\left|\phi\right\rangle_{n} \right. \nonumber \\ &+ &  \left.\left|\xi_{k}\right\rangle^{-}_{a12..n-1}\iota\sigma_{y}\left|\phi\right\rangle_{n}\right]
\end{eqnarray}
Where $\left|\phi\right\rangle_{n}=[\alpha\left|0\right\rangle +\beta\left|1\right\rangle]_{n}~,~\alpha^{2}+\beta^{2}=1 $.\\
Eq. (8) clearly shows that the teleportation protocol is always successful with equal
probability of 1/4 for the four different measurement outcomes of Alice. Therefore, Bob can
always recover the original state by performing single qubit unitary transformations on the
state of his qubit, once he receives the two bit classical message from Alice regarding her
measurement outcome.
\section{Analysis of the efficiency of W-type states in teleportation process}
We have shown that the N-particle W-type state can be successfully used as an optimal
resource for efficient teleportation. The successful completion of teleportation protocol
depends on the availability of experimental set up to perform and distinguish multiqubit
measurements. It is evident that with the present experimental techniques, one can perform and
distinguish different Bell measurements. Therefore, we analyze the efficacy of our states for a protocol where two users
want to create an efficient bi-partite entangled channel between them using the partially entangled four qubit W state
$\left|\Psi_{k}\right\rangle_{1234}$. For this, we assume that Alice initially has a two-qubit
entangled state $\left|\phi\right\rangle_{ab}=\left[\alpha\left|00\right\rangle+\beta\left|11\right\rangle\right]_{ab}~,~\alpha^{2}+\beta^{2}=1$ in addition to the shared W-type entangled state
\begin{eqnarray}
\left|\Psi_{k}\right\rangle_{1234} &=&
\frac{1}{2\sqrt{k+1}}\left[\left|1000\right\rangle+\sqrt{k}\left|0100\right\rangle\right.
\nonumber \\ &+&\left. \sqrt{k+1}\left|0010\right\rangle\right. +
\left.\sqrt{2k+2}\left|0001\right\rangle\right]_{1234}
\end{eqnarray}
with Bob such that qubits $1, 2$ and $3$ are with Alice and qubit $4$ is with Bob. In order to share a bi-partite entanglement with Bob, Alice needs to perform Bell measurements
\begin{eqnarray}
\label{Bell}
\left| \phi\right\rangle^{\pm} &= &\frac{1}{\sqrt{2}} \left[\left| 00 \right\rangle \pm \left| 11 \right\rangle
\right],   \nonumber \\
\left| \psi\right\rangle^{\pm}  &=& \frac{1}{\sqrt{2}} \left[\left| 01 \right\rangle \pm \left| 10 \right\rangle
\right]
\end{eqnarray}
on her qubits. There are different combinations in which Alice can perform these Bell measurements to achieve the required two qubit entanglement. We have examined all possible combinations and measurement outcomes, and here we will discuss only three optimal cases where the concurrence of finally shared two-qubit entangled state is optimal and efficient. We now proceed to analyze the efficacy of the protocol in terms of the concurrence of the final entangled state.

\begin{description}
\item[$\bullet$ Case:1] In the first case, Alice's measurement outcomes are
$\left|\phi^{+}\right\rangle_{b1}$ and $\left|\phi^{+}\right\rangle_{23}$. Therefore, the joint state of two qubits shared between Alice and Bob can be represented as
\begin{eqnarray}
\left|\psi\right\rangle_{a4} &=& \frac{1}{\sqrt{(2k+2)\alpha^{2}+\beta^{2}}}\left[\sqrt{2k+2}\alpha \right. \nonumber \\ && \left.\left|01\right\rangle_{a4} + \beta\left|10\right\rangle_{a4}\right]
\end{eqnarray}
The concurrence of $\left|\psi\right\rangle_{a4}$ is
\begin{equation}
C^{(1)}_{4} = \frac{2\alpha\sqrt{1-\alpha^{2}}\sqrt{2k+2}}{(2k+1)\alpha^{2}+1}
\end{equation}
Where subscript of $C$ represents number of qubit and superscript represent different cases.\\ Eq. (12) clearly demonstrates that for any given real positive number $k$, if $|\alpha|^{2}$ is varied from $0$ to $1$ then concurrence first increases and then decreases to a minimum value. Interestingly, for $\alpha^{2}=\frac{1}{(2k+3)}$ concurrence of the shared entangled state is unity i.e. Alice and Bob can share a maximally entangled state. It is a interesting result since Alice and Bob initially started in a partially entangled state but by performing Bell state measurements they created a bi-partite maximum entanglement between them. The finally shared state, thus, can be used for various information processing protocols. This can be really useful in scenarios where the users in a communication protocol only have access to partially entangled multiqubit states. Further, the analysis presented here not only allows the users to create maximum entanglement but also releases the constraints on the experimental set up to perform and distinguish multiqubit measurements. The price one pays to achieve the maximum entanglement are two standard Bell measurements. Nevertheless, once the users achieve maximum entanglement, the state can be used for various efficient and optimal applications in quantum information and computation. \\
\item[$\bullet$ Case:2] In the second case, Alice's measurement outcomes are $\left|\phi^{+}\right\rangle_{b2}$ and $\left|\phi^{+}\right\rangle_{13}$. Hence the shared bipartite state and concurrence of this state can be given by
\begin{eqnarray}
\left|\psi\right\rangle_{a4} &=& \frac{1}{\sqrt{(2k+2)\alpha^{2}+k\beta^{2}}}\left[\sqrt{2k+2}\alpha \right. \nonumber \\ & &  \left. \left|01\right\rangle_{a4} +\sqrt{k} \beta\left|10\right\rangle_{a4}\right]
\end{eqnarray}
and
\begin{equation}
C^{(2)}_{4} = \frac{2\alpha\sqrt{1-\alpha^{2}}\sqrt{2k+2}\sqrt{k}}{(k+2)\alpha^{2}+k},
\end{equation}
respectively. Similar to the first case discussed above, the concurrence of the shared state first increases; attains the maximum value and then decreases to 0 for any $k$ and $0<\alpha\leq1$. Further, for $\alpha^{2}=\frac{k}{(3k+2)}$ concurrence of the shared state is unity.   \\
\item[$\bullet$ Case:3] The third case provides another interesting observation that for Alice's measurement outcomes are $\left|\phi^{+}\right\rangle_{b3}$ and $\left|\phi^{+}\right\rangle_{12}$, the concurrence of shared bipartite state is independent of the parameter $k$. In this scenario, the shared bipartite state and its concurrence is represented as
\begin{eqnarray}
\left|\psi\right\rangle_{a4} &=&
\frac{1}{\sqrt{(2k+2)\alpha^{2}+(k+1)\beta^{2}}}\left[\right.
\nonumber \\ & &  \left.\sqrt{2k+2}\alpha
\left|01\right\rangle_{a4} \right. +  \left.\sqrt{k+1}
\beta\left|10\right\rangle_{a4}\right]
\end{eqnarray}
and
\begin{equation}
C^{(3)}_{4} = \frac{2\sqrt{2}\alpha\sqrt{1-\alpha^{2}}}{\alpha^{2}+1},
\end{equation}
respectively. The concurrence given in Eq. (16) attains its maximum value i.e. unity for $\alpha^{2}=\frac{1}{3}$. \\
\item[$\bullet$ Case:4] The fourth case i.e. when Alice's measurement outcomes are $\left|\phi^{+}\right\rangle_{a1}$ and $\left|\phi^{+}\right\rangle_{b2}$, also provides another interesting observation such that the concurrence of shared bipartite state is independent of the parameter $k$ and $\alpha$. In this scenario, the shared bipartite state and its concurrence is represented as
\begin{eqnarray}
\left|\psi\right\rangle_{34} &=& \frac{1}{3n+3}\left[\sqrt{2k+2} \left|01\right\rangle_{34} \right. \nonumber \\ &+&  \left.\sqrt{k+1} \left|10\right\rangle_{34}\right]
\end{eqnarray}
and
\begin{equation}
C^{(4)}_{4} = \frac{2\sqrt{2}}{3}
\end{equation}
respectively. The concurrence given in Eq. (18) does not depend on the input state. \\
\end{description}
Fig. (1) compares the first three cases above to analyze the efficacy of shared bipartite state in terms of concurrence. For $k=1$, concurrence for cases 1 and 2 are same. For large $k$, case 2 and case 3 lead to identical results. Moreover, Fig. (1) also shows a relation between $\alpha$ and combination of Bell measurements to be performed to achieve the optimal concurrence.   \par
\begin{figure*}[t]
\centering
\setlength\fboxsep{0pt}
\setlength\fboxrule{0.25pt}
\fbox{\includegraphics[width=\textwidth]{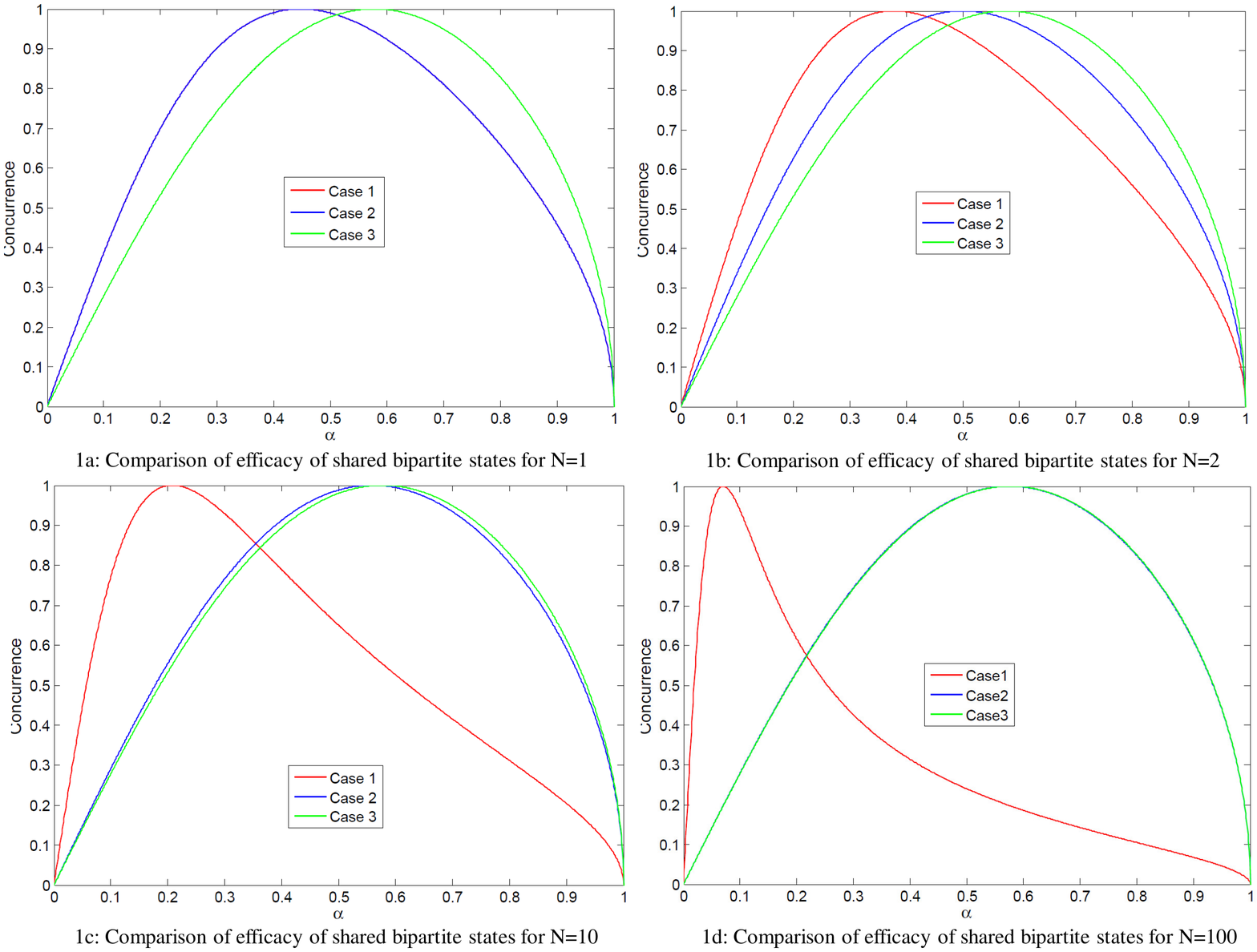}}
\caption{Comparison of efficacy of shared bipartite states in three optimal cases}
\end{figure*}
A similar calculation for a shared $N$ qubit partially entangled state shows that the concurrence of final states, dependent on input parameters, are\\\\
\lefteqn{C=}
\begin{equation}
 \frac{2\alpha\beta\sqrt{k+r}\sqrt{(N-2)k +\frac{(N-2)(N-3)}{2}+1}}{((N-2)k+\frac{(N-2)(N-3)}{2}+1)\alpha^{2}+ (k+r)\beta^{2}}
\end{equation}
where $r$ is a variable and varies from $0$ to $(N-3)$. Eq. (19) clearly depicts that for $r=(1-k)$, it is clear that the entanglement of the final state shared between Alice and Bob depends on the input state parameters $\alpha$ and $k$.
For $k \to \infty$,  the concurrence is given by
\begin{equation}
C = \frac{2\alpha\sqrt{1-\alpha^{2}}\sqrt{N-2}}{(N-3)\alpha^{2}+1}
\end{equation}
Hence, for a given range of $\alpha$, if $k$ is very large then the W-type state with smaller number of particle is a better resource. \\
Similarly the concurrence of final states, independent of input parameters, are
\begin{equation}
C = \frac{2\sqrt{k+r}\sqrt{(N-2)k+\frac{(N-2)(N-3)}{2}+1}}{((N-1)k+\frac{(N-2)(N-3)}{2}+1+r)},
\end{equation}
where $r$ is a variable and varies from $0$ to $(N-3)$. It is evident from Eq. (21) that for $r=(1-k)$, entanglement of the final state shared between Alice and Bob depends only on  $k$.
For $k \to \infty$,  the concurrence is given by
\begin{equation}
C = \frac{2\sqrt{N-2}}{(N-1)}
\end{equation}
Hence, if $k$ is very large then the W-type state with smaller number of particle is a better resource. \\
In order to analyze the usefulness of four qubit W-type states for such a protocol, we further compare the efficacy of three and four-qubit W-type states as resources in terms of concurrence of the finally shared entangled state. We found an interesting observation that for certain range of $\alpha$, the four qubit W-type states are more efficient resources in comparison to three qubit W-type states for achieving optimal concurrence shared between two users. For this, let us first give the form of three qubit W-type states as
\begin{eqnarray}
\left|\Psi_{k}\right\rangle_{123} &=& \frac{1}{\sqrt{2k+2}}\left[\left|100\right\rangle+\sqrt{k}\left|010\right\rangle\right. \nonumber \\ &+ &  \left. \sqrt{k+1}\left|001\right\rangle\right]_{123}
\end{eqnarray}
Similar to the four-qubit case, there are optimal cases for which the concurrences of finally shared states can be given as
\begin{equation}
C^{(1)}_{3} = \frac{2\alpha\sqrt{(1-\alpha^{2})}\sqrt{k+1}}{(k)\alpha^{2}+1}
\end{equation}
and
\begin{equation}
C^{(2)}_{3} = \frac{2\alpha\sqrt{k(k+1)(1-\alpha^{2})}}{\alpha^{2}+k},
\end{equation}
In above two cases the optimal concurrence of finally shared entangled states is dependent on input state. But similar to four-qubit case, there is a one optimal case in which concurrence of finally shared state is independent on input state.
\begin{equation}
C^{3}_{3} = \frac{2\sqrt{k+1}}{(k+2)}
\end{equation}
respectively. Fig. (2) clearly demonstrates the comparison between the efficiencies of three and four-qubit W states in terms of concurrence of shared bipartite state.
\begin{figure*}[t]
\centering
\setlength\fboxsep{0pt}
\setlength\fboxrule{0.25pt}
\fbox{\includegraphics[width=\textwidth]{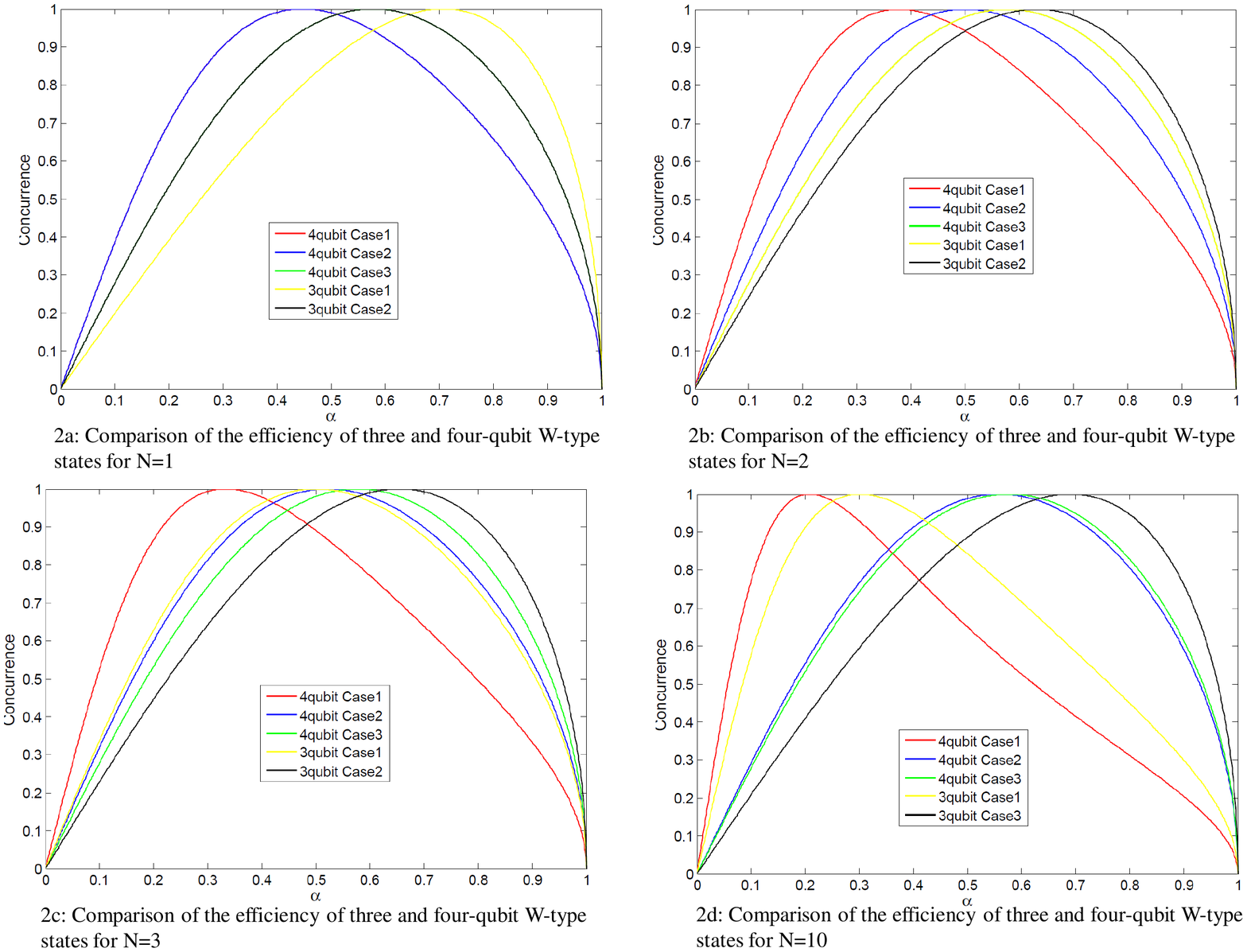}}
\caption{Comparison of the efficiency of three and four-qubit W-type states as resources}
\end{figure*}
Depending on the value of parameter $k$, we identify four different cases; \\
\textbf{Case 1:} For $k=1$ if $0<\alpha^{2}\leq \frac{k(\sqrt{2}-1)}{((k+2)-\sqrt{2})}$ then 4-particle W-type state is a better resource  in comparison to 3-particle W-type state else vice-verse.\\
\textbf{Case 2:} For $k=2$
\begin{description}
 \item[$\bullet$ Range:1]If $0<\alpha^{2}\leq \frac{\sqrt{2}-1}{(2-\sqrt{2})k+1} $ then 4-particle W-type state is a better resource  in comparison to 3-particle W-type state.
 \item[$\bullet$ Range:2]If $\frac{\sqrt{2}-1}{(2-\sqrt{2})k+1}<\alpha^{2}\leq \frac{\sqrt{k+1}-\sqrt{2}}{k\sqrt{2}-\sqrt{k+1}} $ then 3-particle W-type state is a better resource  in comparison to 4-particle W-type state.
 \item[$\bullet$ Range:3]If $\frac{\sqrt{k+1}-\sqrt{2}}{k\sqrt{2}-\sqrt{k+1}}<\alpha^{2}\leq \frac{\sqrt{2}k-\sqrt{k}\sqrt{k+1}}{\sqrt{k}\sqrt{k+1}-\sqrt{2}} $ then 4-particle W-type state is a better resource  in comparison to 3-particle W-type state.
 \item[$\bullet$ Range:4]If $\frac{\sqrt{2}k-\sqrt{k}\sqrt{k+1}}{\sqrt{k}\sqrt{k+1}-\sqrt{2}}< \alpha^{2}<1 $ then 3-particle W-type state is a better resource  in comparison to 4-particle W-type state.
 \end{description}
 \textbf{Case 3:} For $k>2$
\begin{description}
 \item[$\bullet$ Range:1]If $0<\alpha^{2}\leq \frac{\sqrt{2}-1}{(2-\sqrt{2})k+1} $ then 4-particle W-type state is a better resource  in comparison to 3-particle W-type state.
 \item[$\bullet$ Range:2]If $\frac{\sqrt{2}-1}{(2-\sqrt{2})k+1}<\alpha^{2}\leq \frac{k-\sqrt{2k}}{\sqrt{2k}-(k+2)} $ then 3-particle W-type state is a better resource  in comparison to 4-particle W-type state.
 \item[$\bullet$ Range:3]If $\frac{k-\sqrt{2k}}{\sqrt{2k}-(k+2)}<\alpha^{2}\leq \frac{k-\sqrt{k}\sqrt{k+1}}{\sqrt{k}\sqrt{k+1}-(k+2)} $ then 4-particle W-type state is a better resource  in comparison to 3-particle W-type state.
  \item[$\bullet$ Range:4]If $\frac{k-\sqrt{k}\sqrt{k+1}}{\sqrt{k}\sqrt{k+1}-(k+2)}<\alpha^{2}\leq \frac{\sqrt{2}k-\sqrt{k}\sqrt{k+1}}{\sqrt{k}\sqrt{k+1}-\sqrt{2}} $ then 4-particle W-type state is a better resource  in comparison to 3-particle W-type state.
 \item[$\bullet$ Range:5]If $\frac{\sqrt{2}k-\sqrt{k}\sqrt{k+1}}{\sqrt{k}\sqrt{k+1}-\sqrt{2}}< \alpha^{2}<1 $ then 3-particle W-type state is a better resource  in comparison to 4-particle W-type state.
 \end{description}
\textbf{Case 4:} When $k$ is very large
 \begin{description}
 \item[$\bullet$ Range:1]If $0<\alpha^{2}\leq \frac{\sqrt{2}-1}{(2-\sqrt{2})k+1} $ then 4-particle W-type state is a better resource  in comparison to 3-particle W-type state.
 \item[$\bullet$ Range:2]If $\frac{\sqrt{2}-1}{(2-\sqrt{2})k+1}<\alpha^{2}\leq \frac{k-\sqrt{2k}}{\sqrt{2k}-(k+2)} $ then 3-particle W-type state is a better resource  in comparison to 4-particle W-type state.
 \item[$\bullet$ Range:3]If $\frac{k-\sqrt{2k}}{\sqrt{2k}-(k+2)}<\alpha^{2}\leq \frac{\sqrt{2}k-\sqrt{k}\sqrt{k+1}}{\sqrt{k}\sqrt{k+1}-\sqrt{2}} $ then 4-particle W-type state is a better resource  in comparison to 3-particle W-type state.
 \item[$\bullet$ Range:4]If $\frac{\sqrt{2}k-\sqrt{k}\sqrt{k+1}}{\sqrt{k}\sqrt{k+1}-\sqrt{2}}< \alpha^{2}<1 $ then 3-particle W-type state is a better resource  in comparison to 4-particle W-type state.
 \end{description}

Hence, for practical implementation of an efficient bipartite state sharing protocol one can choose  W-type states as resources according to the range of parameters $\alpha$ and $k$.
\section{Superdense coding using W-type states of n-particle system}
Superdense coding deals with efficient information transfer between the users in a communication protocol using a shared entangled resource. We use
\begin{eqnarray}
\left|\eta_{1}\right\rangle_{1234}^{+} &=& \frac{1}{2\sqrt{2}}\left[\left|0100\right\rangle+\left|0010\right\rangle\right. \nonumber \\ &+&  \left.\sqrt{2}\left|0001\right\rangle + 2\left|1000\right\rangle\right]_{1234}
\end{eqnarray}
as a shared resource for superdense coding protocol between Alice and Bob such that the first qubit is with Alice and rest of the qubits are with Bob. In order to communicate the classical message to Bob, Alice first encodes her message using one of the four single qubit operations ${I,\sigma_{x},\sigma_{y},\sigma_{z}}$ on her qubit 1. The four operations map the originally shared state between Alice and Bob to four otrhogonal states
\begin{eqnarray}
\left(\sigma_{x} \otimes I \otimes I \otimes I \right) \left|\eta_{1}\right\rangle_{1234}^{+} &=&  \left|\xi_{1}\right\rangle_{1234}^{+} \nonumber \\
\left(\sigma_{z} \otimes I \otimes I \otimes I \right)\left|\eta_{1}\right\rangle_{1234}^{+} &=&  \left|\eta_{1}\right\rangle_{1234}^{-}  \nonumber \\
\left(i\sigma_{y} \otimes I \otimes I \otimes I\right)\left|\eta_{1}\right\rangle_{1234}^{+} &=&  \left|\xi_{1}\right\rangle_{1234}^{-} \nonumber \\
\left(I \otimes I \otimes I \otimes I\right)\left|\eta_{1}\right\rangle_{1234}^{+} &=& \left|\eta_{1}\right\rangle_{1234}^{+}
\end{eqnarray}
Thus, in principle, Alice can prepare four distinct messages for Bob by locally manipulating her qubit. Once Alice encodes the message, she sends her qubit to Bob. In order to distinguish between the messages sent by Alice, Bob can always perform an appropriate joint measurement on the state of four qubits. Hence, Bob will always be able to distinguish between the four messages produced by Alice. The protocol is optimal as by locally manipulating her one qubit, Alice can transmit two bits of classical message to Bob. \par
We now proceed to demonstrate optimal dense coding protocol using our $N-$particle W-type state
\begin{eqnarray}
\lefteqn{\left|\eta_{k}\right\rangle^{+}_{12...N} =} && \nonumber \\ && \frac{1}{\sqrt{(N-2)(2k+N-3)+2}}\left[\left|010...N\right\rangle\right. \nonumber \\ &+&\left.\sqrt{k}\left|001...N\right\rangle + \sqrt{k+1}\left|0001...N\right\rangle \right. \nonumber \\ &+&\left. .....\sqrt{k+(N-3)}\left|000...1\right\rangle \right. \nonumber \\ &+& \left.\sqrt{(N-2)k+\frac{(N-2)(N-3)}{2}+1}\right. \nonumber \\ & &\left.\left|100...0\right\rangle\right]_{12...N} \nonumber \\
\end{eqnarray}
where qubit 1 is with Alice and rest of the qubits are with Bob. Similar to the four particle case, Alice can produce four distinct messages for Bob using single qubit unitary transformations ${I,\sigma_{x},\sigma_{y},\sigma_{Z}}$ such that
\begin{eqnarray}
\left(I \otimes I \otimes I \otimes I\right)\left|\eta_{k}\right\rangle^{+}_{12...N} &=& \left|\eta_{k}\right\rangle^{+}_{12...N} \nonumber \\
\left(\sigma_{x} \otimes I \otimes I \otimes I \right)\left|\eta_{k}\right\rangle^{+}_{12...N} &=& \left|\xi_{k}\right\rangle^{+}_{12...N} \nonumber \\
\left(\sigma_{z} \otimes I \otimes I \otimes I\right)\left|\eta_{k}\right\rangle^{+}_{12...N} &=& \left|\eta_{k}\right\rangle^{-}_{12...N} \nonumber \\
\left(i\sigma_{y} \otimes I \otimes I \otimes I\right)\left|\eta_{k}\right\rangle^{+}_{12...N} &=& \left|\xi_{k}\right\rangle^{-}_{12...N}
\end{eqnarray}
Therefore our $N$-particle W-type state can also be used for optimal super dense coding protocol.
\section{Conclusion}
We have analyzed a class of partially entangled four-qubit W-type states for efficient quantum
information processing tasks. Although performing and distinguishing multiqubit measurements is an uphill task, nevertheless, our states can be used for deterministic teleportation with unit fidelity. In order to demonstrate the practical utility of such states, we have discussed and compared the efficiency of three and four qubit W-type states for sharing optimal bipartite entanglement between two users. Our results will be of high importance in situations where users only have access to partially entangled states and would like to establish optimal bipartite entanglement for efficient and deterministic information processing.  \par
The analytical relations between the range of state parameters, and optimal concurrence of the finally shared state is also
obtained allowing one to decide when to use a three or four qubit W-type states for a particular
protocol. We have also shown that our states can be used for optimal dense coding as well. The protocols have also been generalized for the case of $N$ qubits.

\end{document}